\documentclass[runningheads]{svmult}

\usepackage{makeidx}   
\usepackage{graphicx}  
\usepackage{subeqnar}  
\usepackage{multicol}  
\usepackage{physprbb}  
\makeindex             

\begin{document}
\title*{Modeling the faint Radio Population: \\
the nanoJy radio sky}
\toctitle{Modeling the faint Radio Population:
\protect\newline The nanoJy radio Sky}
%
%
\titlerunning{The nanoJy radio Sky}
%
\author{Isabella Prandoni\inst{1}
\and Hans R. de Ruiter\inst{2}
\and Paola Parma\inst{1}}
\authorrunning{I. Prandoni et al.}
%
%
\institute{Istituto di Radioastronomia - INAF - Bologna, Italy 
\and Osservatorio Astronomico di Bologna - INAF - Bologna, Italy}

\maketitle              

\begin{abstract}
The apparent change in the composition of the parent optical objects of radio 
sources around 1 mJy (at 1.4 GHz) is now well established, although there is 
still some debate about the relative importance of classical radio galaxies 
and star-forming galaxies at sub-mJy levels (see e.g. Gruppioni et al. 1999, 
MNRAS, 304, 199; Prandoni et al. 2001b, A\&A, 369, 787). It is clear, however, 
that at $\mu$Jy levels star-forming galaxies are dominant (see Fomalont et al. 
1997, ApJ, 475, L5; Haarsma et al. 2000, ApJ, 544, 641).\\
Does this mean that SKA will basically tell us more about the history of 
star formation than about the space density (and its cosmological evolution) 
of active galactic nuclei? \\
Using current best estimates of luminosity functions (and their evolution) of 
various classes of objects, we show that the increasing dominance of 
star-forming galaxies below 
1 mJy is a natural consequence of the different luminosity functions, but that 
this does not at all mean that star-forming galaxies do necessarily dominate 
at all sub-mJy flux levels and all redshifts.
\end{abstract}

\section{Models and Comparison Samples}
In order to have an idea on the types of objects we can expect at flux levels 
accessible with SKA we have modeled the main classes of sources detected 
at mJy and sub-mJy levels: steep AGNs (Radio Galaxies) modeled 
following Dunlop \& Peacock (1990, MNRAS, 247, 19); flat AGNs
(Sy1 and QSO), for which we have assumed the quasar optical LF and evolution
(Boyle et al 1988, MNRAS, 235, 935; 1991, ASP Conf. Ser. 21, p.~191; Schmidt 
et al 1995, AJ, 109, 473); the star-forming galaxies (RLF from Sadler et al 
2002, MNRAS, 329, 227) composed by a fraction (assumed 50\%) of non-evolving 
normal spirals and a fraction (50\%) of evolving starburst galaxies 
($L\sim (1+z)^3$). Passive optical evolution has been assumed whenever 
necessary (Poggianti 1997, A\&AS, 122, 399). \\
A number of available surveys at the mJy, sub-mJy and µJy level can provide 
important boundary conditions to any modelling of the radio sky.
The radio counts are constrained by using all the samples available in the 
literature, while we focused on samples with optical spectroscopy follow-up 
to get constraints on the redshift and magnitude distributions of the sources. 
In particular we refer to the following samples: FIRST (Magliocchetti et al. 
2000, MNRAS, 318, 1047), ATESP-EIS (Prandoni et al. 2001b), PDF (Phoenix Deep 
Field, Georgakakis et al. 1999, MNRAS, 306, 708), MF (Marano Field, Gruppioni 
et al. 1999), B93 (sample collection studied by Benn et al. 1993, MNRAS, 263, 
98), H00 (collection studied by Haarsma et al. 2000). \\
The models used here provide a good fit to the observed number counts along 
the entire flux range spanned by the counts ($40$ $\mu$Jy ­- 1 Jy) and can 
reproduce the total number of sources in the comparison samples within a 
factor of 2. The models can trace with good accuracy both the magnitude and 
the redshift distributions of the sources in the given samples.

\section{The Composition of the nanoJy Radio Sky}
\begin{figure}[t]
\begin{center}
\includegraphics[width=.65\textwidth]{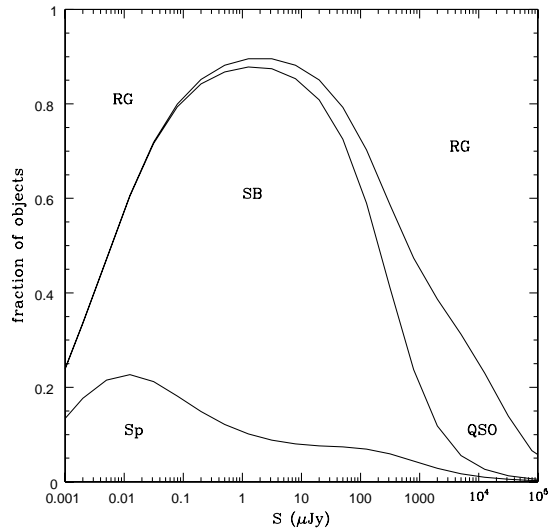}
\end{center}
\caption[]{Summed contribution of different kinds of sources as a 
function of flux. Regions separated by solid lines correspond to: 
steep-spectrum sources (RG), flat-spectrum sources (QSO), normal non-evolving 
spirals (Sp) and starbursts (SB).}
\label{eps1}
\end{figure}

The models above have been used to simulate the radio sky at fainter
flux levels than reached by the current surveys. The composition of the 
radio sky changes with flux as shown in Figure~\ref{eps1}. 
The figure clearly shows that radio galaxies, which dominate (together with 
QSO) the mJy population, reappear in large proportions going to nanoJy 
levels ($>50\%$ at $S<10$ nJy)!\\
On the other hand, starburst galaxies and their evolution can be suitably 
studied with less sensitive surveys (e.g. $S>10 - 100$ nJy). 
The other main population at nanoJy level is represented by 
non-evolving spirals, whose contribution shows a bump (mainly due to $z>1$ 
galaxies) in the range $1 < S < 100$ nJy.\\
This work demonstrates that nuclear activity could be important at
nanoJy flux levels. Deeper data are strongly needed to better constrain the 
models and provide more reliable simulations. This kind of analysis can 
provide very useful constraints to the design of SKA.

%

\end{document}